\documentclass[aps,superscriptaddress,10pt,nofootinbib,notitlepage,twocolumn,prd]{revtex4-2}

\usepackage[normalem]{ulem}
\usepackage[latin1]{inputenc}
\usepackage{graphicx}
\graphicspath{{Figures/}}
\usepackage{float}
\usepackage{latexsym}
\usepackage{graphicx}
\usepackage{amssymb}
\usepackage{amsmath}
\usepackage{amsfonts}
\usepackage[colorlinks=true,citecolor=blue,hyperfootnotes=false]{hyperref}
\usepackage[dvipsnames]{xcolor}
\usepackage{cool}
\usepackage{dcolumn}
\usepackage{textcomp}
\usepackage{xfrac}
\usepackage{slashed}
\usepackage{multirow}
\usepackage{orcidlink}

\def\be{\begin{equation}}
\def\ee{\end{equation}}
\def\figs/B{B}
\def\bea{\begin{eqnarray}}
\def\eea{\end{eqnarray}}
\def\bg{\begin{eqnarray}}
\def\nd{\end{eqnarray}}

\def\cos{{\rm cos}}

\def\ln{{\rm ln}}

\def\mpl{{M_{\rm Pl}}}
\def\dd{{\rm d}}

\begin{document}

\title{Natural Inflation with Exponentially Small Tensor-To-Scalar Ratio}

\author{Dario L.~Lorenzoni \orcidlink{0000-0001-9832-8183}}
\email{lorenzod@myumanitoba.ca}
\affiliation{Department of Physics \& Astronomy, University of Manitoba, Winnipeg, Manitoba R3T 2N2, Canada}

\author{David I.~Kaiser \orcidlink{0000-0002-5054-6744}}
\email{dikaiser@mit.edu}
\affiliation{Department of Physics,
Massachusetts Institute of Technology, Cambridge, MA 02139, USA}

\author{Evan McDonough \orcidlink{0000-0002-2130-3903}}
\email{e.mcdonough@uwinnipeg.ca}
\affiliation{Department of Physics, University of Winnipeg, Winnipeg MB, R3B 2E9, Canada}

\begin{abstract}
    We demonstrate that ``natural inflation,'' also known as ``axion inflation,'' can be compatible with {\it Planck} 2018 measurements of the cosmic microwave background, while predicting an exponentially small tensor-to-scalar ratio, e.g., $r\sim 10^{-15}$. The strong suppression of $r$ arises from dynamics of the radial component of the complex scalar field, whose phase is the axion. Such tiny values of $r$ remain well below the threshold for detection by CMB-S4 or Simons Observatory B-mode searches. The model is testable with the running $\alpha_s$ of the spectral index, which is within reach of next-generation CMB and large-scale structure experiments, motivating the running as a primary science goal for future  experiments.
\end{abstract}

\maketitle

\textbf{\textit{Introduction.}} 
Cosmic inflation is the leading description of the very early Universe. It provides a causal mechanism for the generation of large-scale structure of the Universe, observed both by large-scale structure surveys and as anisotropies in the cosmic microwave background (CMB). (For reviews, see, e.g., Refs.~\cite{Lyth:2009imm,Martin:2013tda,Guth:2013sya,Baumann:2022mni}.) The discovery of acoustic peaks in the CMB handed inflation its first decisive victory over then-rival  cosmic strings \cite{Dodelson:2003ip,Dvorkin:2011aj,Urrestilla:2011gr}. Subsequent measurements by the Wilkinson Microwave Anisotropy Probe~\cite{WMAP:2012nax} and {\it Planck}~\cite{Planck:2018jri} collaborations have further bolstered the case for inflation, for example, with measurements of the spectral index of the primordial power spectrum in agreement with predictions from various inflation models.  

Much attention has been paid to the possibility that next-generation CMB experiments, such as CMB-S4 \cite{CMB-S4:2016ple} and the Simons Observatory \cite{Ade:2018sbj}, could detect gravitational waves produced by inflation, which have an amplitude parametrized by the tensor-to-scalar ratio $r$.  Yet predictions for the tensor-to-scalar ratio remain strongly model-dependent. In this work, we consider a well-motivated model of inflation that predicts a value of $r$ too small to be observed by any conceivable future experiment, finding instead that {\it other} CMB observables, such as improved constraints on the running of the spectral index, would provide concrete tests of such models. (See also Refs.~\cite{Adshead:2010mc,Hardwick:2018zry,Easther:2021rdg,Bahr-Kalus:2022prj,Martin:2024nlo,Munoz:2016owz}.)

Many models of inflation have been proposed \cite{Martin:2013tda,Martin:2024qnn}. A particularly well motivated example is that of ``natural inflation'' \cite{Freese:1990rb}, also known as ``axion inflation.'' This model builds on the axion model of particle physics, initially proposed as a solution to the strong $CP$ problem \cite{Peccei:1977hh,Wilczek:1977pj,Weinberg:1977ma}, later as a candidate for cold dark matter \cite{Preskill:1982cy,Abbott:1982af,Dine:1982ah}, and yet later discovered to be ubiquitous in both string theory \cite{Svrcek:2006yi,Arvanitaki:2009fg,Cicoli:2012sz} and field theory \cite{Maleknejad:2022gyf,Alexander:2023wgk,Alexander:2024nvi}. It is therefore only ``natural'' to consider an axion-like particle as an inflaton candidate.

However, the predictions of natural inflation as originally formulated in Ref.~\cite{Freese:1990rb} are strongly disfavored by data \cite{Martin:2013tda,Planck:2018jri}. Upon fixing model parameters to yield a prediction for the scalar spectral index $n_s$ within the range favored by data, the predicted tensor-to-scalar ratio becomes $r\sim 0.1$, well in excess of the current observational upper bound $r < 0.036$ \cite{BICEP:2021xfz}. Several works \cite{Achucarro:2015caa,McDonough:2020gmn,Alam:2024krt,Montefalcone:2022jfw,Salvio:2021lka,Salvio:2023cry} have considered the possibility that {\it multifield} inflationary dynamics can bring natural inflation into agreement with current observations.\footnote{Modified gravity can also be used as a means to the same end \cite{Racioppi:2024zva,Salvio:2019wcp,Salvio:2022mld}. Modifications to the postinflationary Universe, i.e., the  $\Lambda$CDM model, have also been considered \cite{RoyChoudhury:2022rva,Bostan:2023ped}.} In what follows, we extend this to natural inflation consistent with a future non-observation of $r$, by analyzing a regime that predicts an exponentially small tensor-to-scalar ratio $r$, namely, $r\sim 10^{-n}$ with $n\gg 1$.\\

\textbf{\textit{Multifield Dynamics in Natural Inflation.}} 
Our starting point is natural inflation \cite{Freese:1990rb}  in its full form, namely, the theory of a spontaneously broken global U(1) symmetry, with action \cite{McDonough:2020gmn}
\begin{eqnarray}\label{eq:model}
    S= && \int \dd ^4 x  \sqrt{-g}\,
    \bigg[  \frac{1}{2}M^2R  - \frac{1}{2} \vert \partial \Phi \vert^2 \\
    && - \frac{\lambda}{4}(|\Phi|^2-v^2)^2 - \Lambda^4(1-\cos \vartheta) + \frac{1}{2} \xi |\Phi|^2 R \bigg] \nonumber \,,
\end{eqnarray}
where $\Phi\equiv \varphi e^{i \vartheta}$; both $\varphi$ and $\vartheta$ are real-valued scalar fields. As required by consistent renormalization in curved spacetime, we include a nonminimal coupling $\xi |\Phi|^2 R$ \cite{Chernikov:1968zm,Callan:1970ze,Bunch:1980br,Bunch:1980bs,Birrell:1982ix,Odintsov:1990mt,Buchbinder:1992rb,Parker:2009uva,Markkanen:2013nwa}. In the spirit of effective field theory, we consider the dimensionless parameter $\xi \simeq {\cal O}(1)$ to be fixed by comparisons with observations. The potential energy includes contributions from two sources: a Higgs-like symmetry-breaking potential and a conventional axion potential for the phase $\vartheta$, associated with a nonperturbative breaking of the continuous axion shift symmetry to a periodic shift symmetry.

In the vacuum of the theory, with $\langle |\Phi| \rangle=v$, this model simplifies to the usual model of axion inflation with axion decay constant $f_a = v $, and the gravitational action reduces to the Einstein-Hilbert action with the identification that $\mpl^2 = M^2 + \xi v^2$, where $\mpl \equiv 1 / \sqrt{ 8 \pi G}$ is the reduced Planck mass. In this limit, this model can realize  natural inflation \cite{Freese:1990rb}. The latter is in significant tension with observations and is essentially ruled out by {\it Planck} 2018 CMB data \cite{Martin:2013tda,Planck:2018jri}.

However, the radial (``Higgs") mode $\varphi$ need not be in its vacuum state in the very early Universe. If $\varphi$ is instead displaced from its minimum, multifield inflation can ensue, wherein both $\varphi$ and $\vartheta$ are dynamical and contribute to the expansion history of the Universe.

The background evolution of the model in Eq.~\eqref{eq:model} can most easily be understood by rescaling the spacetime metric to make the gravitational action take the standard Einstein-Hilbert form, via the transformation $g_{\mu \nu} \rightarrow [ \mpl^2  /({M^2 + \xi \varphi^2}) ] {g}_{\mu \nu}$ \cite{Kaiser:2010ps,Abedi:2014mka}. This rescales the potential terms in Eq.~\eqref{eq:model}  as $V (\varphi,\vartheta)\rightarrow \mpl^4 V (\varphi,\vartheta) / (M^2 + \xi \varphi^2)^{2}$, and generates a noncanonical field-space metric ${\cal G}$ with nonvanishing components:
\begin{equation}
    {\cal G}_{\varphi \varphi} = \frac{ \mpl^2}{M^2 + \xi \varphi^2} \left( 1 +  \frac{ 6 \xi^2 \varphi^2}{M^2 + \xi \varphi^2} \right), \> {\cal G}_{\vartheta\vartheta} = \frac{ \mpl^2 \,\varphi^2}{M^2 + \xi \varphi^2} \, . 
\end{equation}
The equations of motion for the fields $\phi^I = \{ \varphi, \vartheta \}$ take the form ${\cal D}_t \dot{\phi}^I + 3 H \dot{\phi}^I + {\cal G}^{IK} V_{, K} = 0$, where the covariant directional derivative ${\cal D}_t$ acting on a field-space vector $A^I$ is defined via ${\cal D}_t A^I \equiv \dot{A}^I + \dot{\phi}^J \Gamma^I_{JK} A^K$ and the field-space Christoffel symbols are evaluated in terms of ${\cal G}_{IJ}$ and its derivatives. The Friedmann equation may be written as $H^2 = [ \frac{1}{2} \dot{\sigma}^2 + V ] / (3 \mpl^2)$, where $\dot{\sigma} \equiv [ {\cal G}_{IJ} \dot{\phi}^I \dot{\phi}^J ]^{1/2}$ \cite{Kaiser:2012ak}.

From this one can appreciate the hallmark features of multifield natural inflation \cite{McDonough:2020gmn}: 
    (1) The model can realize inflation along the radial ($\varphi$) direction. At large values of $\sqrt{\xi} \varphi/M$ (not necessarily large $\sqrt{\xi}$ or $\varphi$), the $\varphi$ sector of the theory reduces to Higgs inflation \cite{Bezrukov:2007ep,Bezrukov:2010jz,Greenwood:2012aj,Rubio:2018ogq}, wherein the potential energy is exponentially stretched, allowing for an extended period of inflation along the radial direction. 
    (2) The axion decay constant is {\it dynamical}. Defined by the axion kinetic term, the decay constant is given by
    \begin{equation}
        f_a = \frac{ \varphi }{M^2 + \xi \varphi^2} \mpl \, .
    \end{equation}
    (3) The axion potential energy and hence its mass is {\it suppressed} at large values of the radial field as
    \begin{equation}\label{eq:Vaxion}
        V_{\rm axion} = \frac{ \mpl^4\Lambda^4 }{( M^2 + \xi \varphi^2)^2}\left( 1 - \cos \vartheta\right) \, .
    \end{equation}
    This naturally makes $\vartheta$ a subdominant component in an early phase of $\varphi$-inflation.
These features combine to allow a multi-phase inflation model, wherein the axion is initially relegated to a spectator field and only becomes important to the dynamics at later stages of inflation \cite{McDonough:2020gmn}. 

The phases of inflation can be understood by defining a pseudoscalar turn rate $\omega$. The unit vector $\hat{\sigma}^I \equiv \dot{\phi}^I / \dot{\sigma}$ indicates the (instantaneous) direction in field space along which the system evolves \cite{Kaiser:2012ak}, in terms of which one may define the turn-rate vector $\omega^I \equiv {\cal D}_t \hat{\sigma}^I$ and the pseudoscalar turn rate $\omega \equiv \epsilon_{IJ} \, \hat{\sigma}^I \, \omega^J$ \cite{McDonough:2020gmn}. (Here $\epsilon_{IJ} \equiv [ {\rm det} \, {\cal G}_{IJ} ]^{1/2}  \, \bar{\epsilon}_{IJ}$, where $\bar{\epsilon}_{IJ}$ is the usual Levi-Civita symbol.)
In a flat field space, with radial and angular fields $r$ and $\theta$, the scalar turn rate is simply $\dot{\theta}$ \cite{Gordon:2000hv}.\\

\textbf{\textit{Cosmological Perturbations.}} 
Perturbations in this model can be decomposed  into an adiabatic (curvature) perturbation and an isocurvature (entropy) perturbation, corresponding to gauge-invariant fluctuations parallel with and orthogonal to the background fields' field-space trajectory, respectively \cite{Gordon:2000hv,Wands:2007bd,Langlois:2008mn,Peterson:2010np,Achucarro:2010da,Gong:2011uw,Kaiser:2012ak,Gong:2016qmq}. To linear order in fluctuations, the equation of motion for a Fourier mode of the comoving curvature perturbation is given by \cite{McDonough:2020gmn}
\begin{equation}\label{eq:zetaEOM}
    \frac{d}{dt}  \left(\dot {\cal R}_k - 2 \omega {\cal S}_k \right) \\
    + (3 + \delta) H \left( \dot  {\cal R}_k - 2 \omega {\cal S}_k \right)  + \frac{k^2}{a^2}{\cal R}_k = 0 \, ,
\end{equation}
where ${\cal S}$ is the comoving isocurvature perturbation, and $\delta \equiv \dot{\epsilon} / (H \epsilon) = 4 \epsilon - 2 \eta$, where (as usual) $\epsilon \equiv - \dot{H} / H^2$ and $\eta \equiv 2 \epsilon - \dot{\epsilon} / (2 H \epsilon)$. The comoving isocurvature perturbation satisfies 
\begin{equation}
    \ddot{\cal S}_k+ (3+ \delta) H \dot{\cal S}_k + \left(\frac{k^2}{a^2} + \mu^2 _s - 4 \omega^2\right){\cal S}_k = - 2 \omega \dot{\cal R}_k \, ,
\end{equation}
where 
\begin{equation}\label{eq:mus}
    \mu_s ^2 = {\cal M}_{ss} + 3 \omega^2 + H^2(2 \epsilon-\eta)(3+5 \epsilon-\eta)+H^2 \eta \kappa \, ,
\end{equation}
with $\kappa \equiv \dot{\eta}/(\eta H)$. Here ${\cal M}_{ss} \equiv ({\cal G}^{IJ} - \hat{\sigma}^I \hat{\sigma}^J ) {\cal M}_{IJ}$ is the projection of the mass-squared matrix ${\cal M}^I \,_J \equiv {\cal G}^{IK} {\cal D }_J {\cal D}_K V - {\cal R}^{I} \,_{LMJ} \dot{\phi}^L \dot{\phi}^M$ onto the isocurvature direction \cite{Kaiser:2012ak}. 

This system dramatically simplifies on super-Hubble scales: the curvature perturbation is sourced by isocurvature modes,
\begin{equation}\label{eq:dotRS}
    \dot{\cal R}_k= 2 \omega {\cal S}_k \, ,
\end{equation}
while the isocurvature modes evolve with time-dependent mass,
\begin{equation}
    \ddot{\cal S}_k+ (3+ \delta) H \dot{\cal S}_k +  \mu^2 _s {\cal S}_k = 0 \, ,
\end{equation}
where we have used $k^2/a^2 \rightarrow 0$ and $\dot{\cal R}=2 \omega {\cal S}$. As indicated by Eq.~\eqref{eq:dotRS}, even in the long-wavelength limit, isocurvature modes ${\cal S}_k$ can transfer power to adiabatic curvature modes ${\cal R}_k$ whenever the background fields' trajectory undergoes turning, with $\omega \neq 0$ \cite{Gordon:2000hv,Wands:2007bd,Langlois:2008mn,Peterson:2010np,Achucarro:2010da,Gong:2011uw,Kaiser:2012ak,Gong:2016qmq}.

The multifield natural inflation model of Eq.~\eqref{eq:model} is characterized by {\it tachyonic} isocurvature perturbations, namely, $\mu_s ^2 <0$. This arises because, at early times, when the background is dominated by $\varphi$ and $\epsilon, \eta, |\omega|/H \ll 1$, the isocurvature direction is approximately $\vartheta$ and the isocurvature mass is approximately ${\cal M}_{ss} \simeq G^{\vartheta \vartheta} \partial_\vartheta ^2 V_{\rm axion}$ with $V_{\rm axion}$ given by Eq.~\eqref{eq:Vaxion}. As $\varphi$ decreases over the course of inflation, the axion becomes increasingly tachyonic, leading to an efficient growth of modes ${\cal S}_k$ on super-Hubble scales. Meanwhile, the decrease in $\varphi$ also triggers a turn in field space, thereby converting the enhanced isocurvature perturbation into a sourced adiabatic curvature perturbation.  The resulting curvature perturbation can be many orders of magnitude larger than the naive single-field estimate \cite{McDonough:2020gmn}. \\

\textbf{\textit{Example.}} 
To illustrate these dynamics, we consider a fiducial example. We numerically solved for the evolution of the background quantities $\varphi (t), \vartheta (t), H (t)$ as well as the evolution of perturbations ${\cal R}_k (t)$, ${\cal S}_k (t)$, imposing Bunch-Davies initial conditions for the field fluctuations. We also performed an independent check of the numerical results using the software package \texttt{PyTransport}~\cite{Mulryne:2016mzv}. For our fiducial example, parameters are given by
\begin{equation}
\begin{gathered}
    \lambda=1.916\times 10^{-21}\;,\;
    \Lambda=2.252\times 10^{-6} \mpl \;,\\
    v=0.443 \mpl\;,\;
    M=0.141 \mpl\;,\;
    \xi=5
\end{gathered}
\end{equation}
and initial conditions
\begin{equation}
    \varphi_i = 4 \mpl\;,\;\vartheta_i  = \pi(1- 10^{-8})
    \;,\;\dot{\varphi}_i=\dot\vartheta_i=0 \, . 
\end{equation}
Note the significant fine-tuning of the initial condition for $\vartheta$, along the lines of the ``extreme axion'' scenario (see, e.g., Ref.~\cite{Winch:2023qzl}). In the present case, this is a reflection that the desired dynamics, while possible and therefore serving as a proof of principle, are not generic.

\begin{figure}[h!]
    \includegraphics[width=0.47\textwidth]{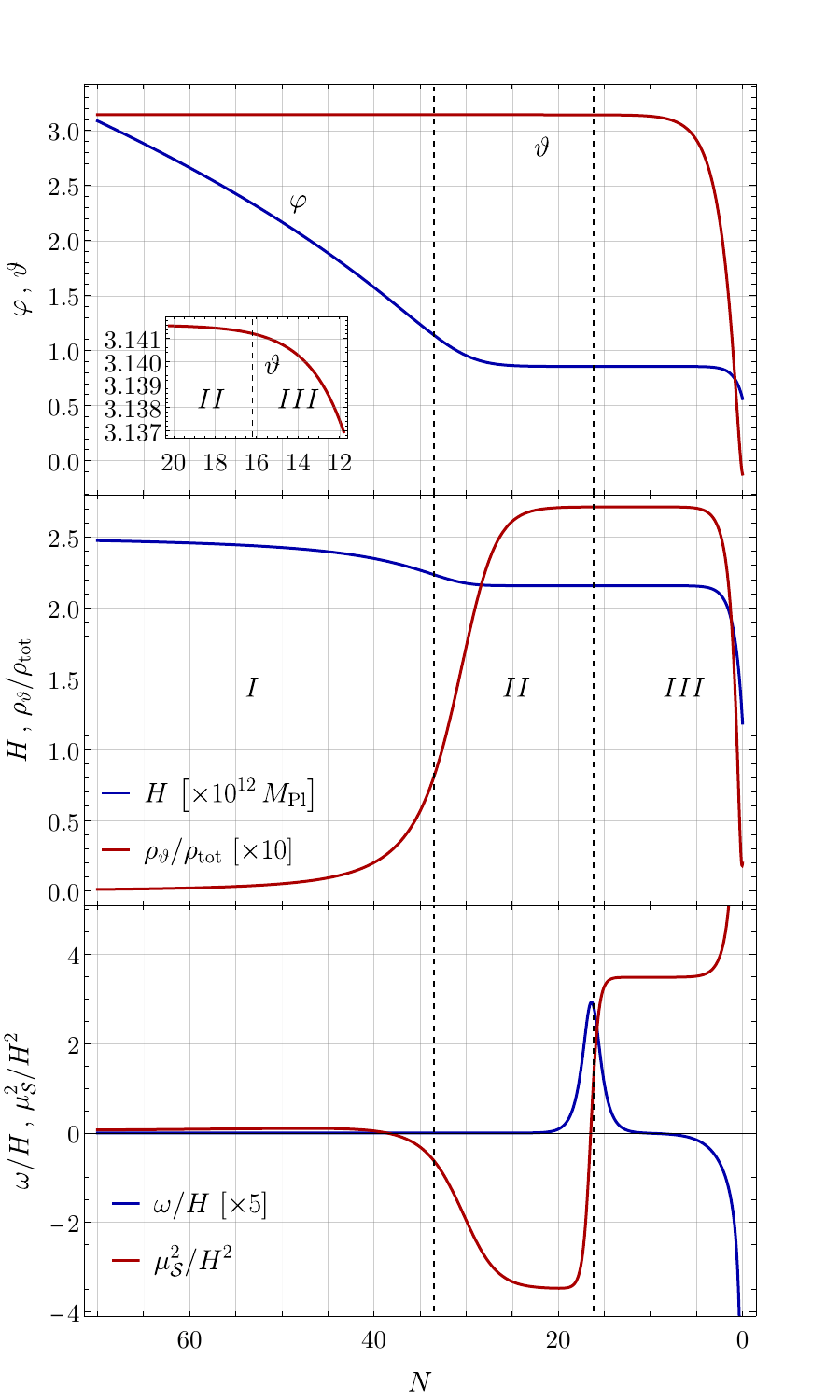}
    \caption{Evolution of the background quantities as functions of the number of efolds $N$ before the end of inflation. The vertical dashed lines delimit the three major inflationary phases, denoted I---III. 
    {\it Top Panel:} Evolution of the fields $\varphi$ [in units of $\mpl$] and $\vartheta$. During phase I, inflation is driven solely by the radial field $\varphi$, while in the latter two phases both fields contribute to the dynamics. The inset shows that $\vartheta$ starts decaying at the boundary between phases II and III, causing a turn in the field-space trajectory.
    {\it Middle Panel:} Evolution of the Hubble parameter $H(t)$ and the fraction of energy density contributed by the axion field, $\rho_\vartheta$.
    {\it Bottom Panel:} Evolution of the turn rate $\omega$ and the isocurvature effective mass $\mu_s$. 
    }
    \label{fig:background}
\end{figure}

Figure~\ref{fig:background} shows the evolution of the background quantities. Note that, for the selected parameters, the model yields low-scale inflation, with $H \sim 10^{-12} \mpl$. There are three distinct phases of the evolution. At early times (phase I), when the dynamics are dominated by the radial field $\varphi$, the turn rate and isocurvature mass are negligible. In phase II, the isocurvature mass-squared becomes {\it negative} while the turn rate remains small. Phase III is then characterized by negligible turning and heavy isocurvature modes, while at the interface between phases II and III, the turn rate briefly becomes large, $\omega / H \sim {\cal O} (1)$, and the isocurvature mass-squared transitions from large and negative to large and positive. The fact that $\mu_s / H > 1$ during phase III suppresses the final amplitude of the long-wavelength modes ${\cal S}_k$.

\begin{figure*}
    \includegraphics[width=0.48\textwidth]{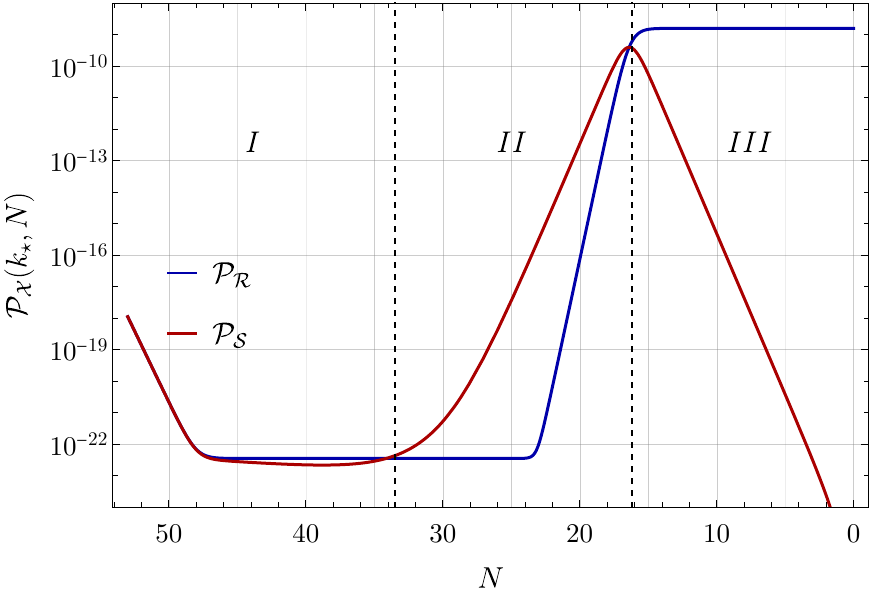}
    \includegraphics[width=0.48\textwidth]{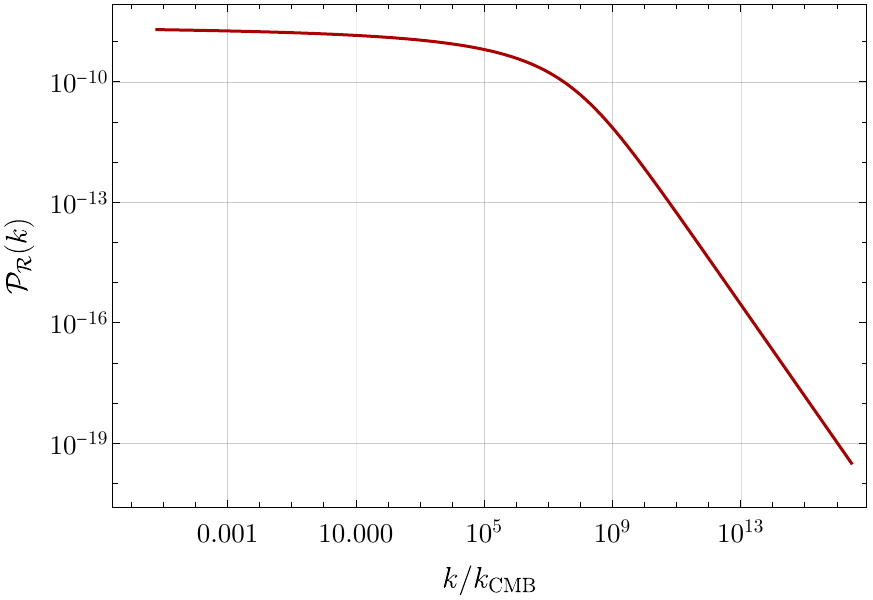}
    \caption{Evolution of primordial perturbations. {\it Left Panel:} Evolution of the dimensionless power spectra for the curvature and isocurvature modes with comoving wavenumber $k_*$, which exit the Hubble radius at $N_*= 49$.
    During phase II of the inflationary evolution, the tachyonic instability of $\mu_s$ enhances the ${\cal S}_k$ mode. Between phases II and III, the field-space turn takes place, and the ${\cal S}_k$ mode transfers power to the ${\cal R}_k$ mode, as in Eq.~\eqref{eq:dotRS}. During phase III, ${\cal R}_k$ is frozen while ${\cal S}_k$ decays due to its large positive mass.
    {\it Right Panel:} Curvature power spectrum ${\cal P}_{\cal R}(k)={\cal P}_{\cal R}(k,N_{\rm end})$. The spectrum is nearly flat at CMB scales (for modes exiting the Hubble radius around $N_*\in \{46.5,51.5\}$) and decreases for modes exiting the Hubble radius at later times, which do not experience any tachyonic instability.}
    \label{fig:R-and-S}                     
\end{figure*}

The left panel of Fig.~\ref{fig:R-and-S} displays the evolution of the dimensionless power spectra ${\cal P}_{\cal X} (k, N) = k^3 | {\cal X}_k (N) |^2 / (2 \pi^2)$ for the curvature and isocurvature perturbations for fixed comoving wavenumber $k_* = 0.05~{\rm Mpc}^{-1}$, corresponding to the CMB pivot scale. As noted below, perturbations with this wavenumber first cross outside the Hubble radius during phase I. Given the low scale of inflation in this scenario, with $H \sim 10^{-12} \mpl$, the power spectra ${\cal P}_{\cal X}$ are exponentially lower during phase I than the COBE normalization, $A_s = 2.1 \times 10^{-9}$. The amplitude of the isocurvature mode ${\cal S}_k$ then grows exponentially during phase II, driven by its tachyonic mass $\mu_s^2 < 0$. As the turn rate $\omega$ rises rapidly around the interface between phases II and III, power is transferred from ${\cal S}_k$ to ${\cal R}_k$, after which $\dot{\cal R}_k \simeq 0$ while the amplitude of ${\cal S}_k$ falls rapidly, since $\mu_s / H > 1$ during phase III. Hence by the end of inflation, we find ${\cal P}_{\cal R} (k_*, N_{\rm end}) = A_s$.

Repeating this calculation for all $k$ that exit the Hubble radius during inflation, we may calculate the primordial power spectrum at the end of inflation as a function of wavenumber, ${\cal P}_{\cal R}(k) = {\cal P}_{\cal R} (k, N_{\rm end})$. This is shown in the right panel of Fig.~\ref{fig:R-and-S}. Note the exponential suppression of modes that exit the Hubble radius after the tachyonic phase for the isocurvature modes has ended, and which therefore do not experience any super-Hubble growth.\\

\begin{figure*}
    \includegraphics[width=0.88\textwidth]{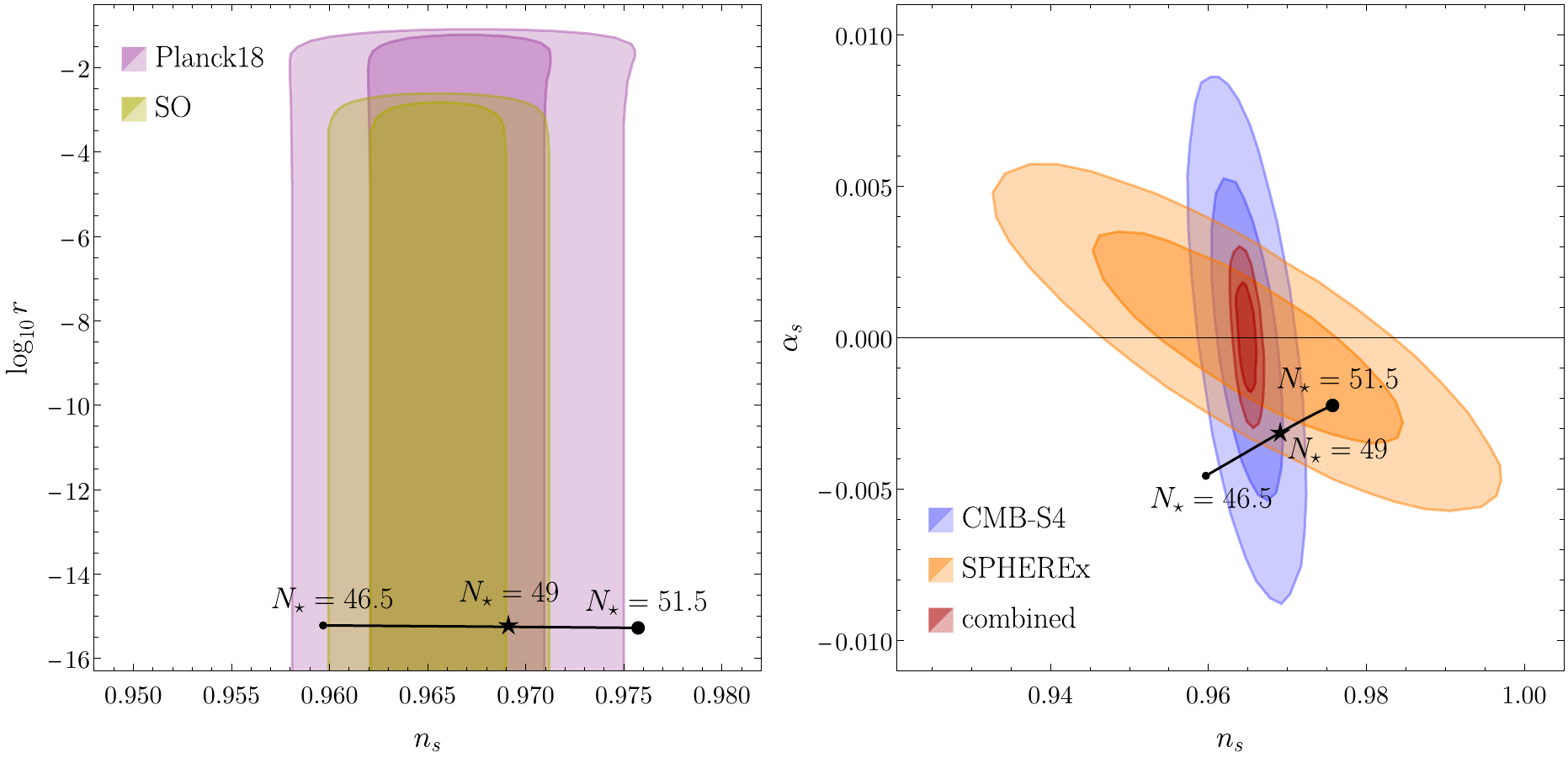}
    \caption{CMB predictions and forecast experimental constraints. {\it Left Panel:} The $n_s$-$r$ plane, with current and forecast constraints shown from {\it Planck} 2018 \cite{Akrami:2018odb} and the Simons Observatory \cite{Ade:2018sbj}. (Image adapted from Ref.~\cite{Kallosh:2019eeu}.) {\it Right Panel:} The $n_s$-$\alpha_s$ plane, with forecast constraints from CMB-S4 \cite{CMB-S4:2016ple} and SPHEREx \cite{SPHEREx:2014bgr}. (Image adapted from Ref.~\cite{Bahr-Kalus:2022prj}.) Superimposed are the predictions for the multifield natural inflation model, which depend on the value of $N_\star$, the time during inflation when perturbations on CMB scales exit the Hubble radius.}
    \label{fig:ns-r}
\end{figure*}

\textbf{\textit{CMB Observables.}} 
We now turn to predictions for observables for this model. To do so, we identify the time of first Hubble crossing of the comoving wavenumber of the CMB pivot scale, $k_*=0.05~{\rm Mpc}^{-1}$, via the standard relation \cite{Dodelson:2003vq,Liddle:2003as}
\begin{eqnarray}
    N_{*} &&\simeq 62 + \frac{1}{4} {\rm ln} \left( \frac{ \rho_*^2}{3 \mpl^6 H^2_{\rm end} } \right) + \frac{1-3 w_{\rm reh}}{12(1+ w_{\rm reh})}   \ln\left(\frac{\rho_{\rm RD}}{\rho_{\rm end}}\right) \nonumber \\
    &&\simeq 49.0 \pm 2.5 \, ,
\end{eqnarray}
where the uncertainty $\pm 2.5$ reflects a duration of reheating $N_{\rm reh} \leq 5$ and an equation of state during reheating within the range $w_{\rm reh} \in \{ - 1/3,+1 \}$. (Reheating in related multifield models with nonminimal couplings has been found to be efficient, with $N_{\rm reh} \leq 5$ across broad regions of parameter space \cite{Ema:2016dny,DeCross:2015uza,DeCross:2016fdz,DeCross:2016cbs,Sfakianakis:2018lzf,Iarygina:2018kee,Nguyen:2019kbm,vandeVis:2020qcp,Iarygina:2020dwe,Bettoni:2021zhq,Ema:2021xhq,Dux:2022kuk,Figueroa:2024asq,Laverda:2023uqv}.) Quantities marked with an asterisk ($*$) are evaluated at the time when $k_* = a (t_*) H(t_*)$ during inflation; quantities denoted ``end'' are evaluated at the end of inflation; and $\rho_{\rm RD}$ is the value of the energy density when the Universe first attains a radiation-dominated equation of state following the end of inflation. The central value $N_{*}=49$ corresponds to instant reheating, or reheating with $w_{\rm reh}=1/3$, whereas $N_*< 49$ ($>49$) implies $w_{\rm reh}< 1/3$ ($> 1/3$).

The exponential enhancement of curvature perturbations as shown in Fig.~\ref{fig:R-and-S} implies an exponential {\it suppression} of the tensor-to-scalar ratio $r$ relative to that at the Hubble crossing:
\begin{equation}\label{eq:radjust1}
    r = \frac{r_*}{1 + (\Delta {\cal R}/{\cal R}_* )^2} \, ,
\end{equation}
where $\Delta {\cal R}$ is the amount of super-Hubble growth of the scalar curvature perturbation. Note that the tensor modes $h_k$ are unaffected by the turn in field space: the equation of motion remains that of single-field inflation, $u_k '' + ( k^2 - a''/a )u_k = 0$, with $u_k \equiv a h_k$ and primes denoting derivatives with respect to conformal time, $d\tau = dt / a$. This equation can be solved in the long wavelength limit by $u_k \propto a$, implying $h_k \simeq {\rm constant}$ \cite{McDonough:2020gmn}. Thus the relative enhancement of scalar curvature perturbations amounts to an overall suppression of the tensor-to-scalar ratio, by the amount given in Eq.~\eqref{eq:radjust1}. For the numerical example of Fig.~\ref{fig:R-and-S}, we find $r= 6 \times 10^{-16}$.

The spectral index of perturbations $n_s$ is also impacted by the growth and transfer of power among the perturbations. While the turn rate $\omega$ acts as a window function for the conversion of isocurvature perturbations into curvature perturbations, the tachyonic instability $\mu_s^2 < 0$ is more effective for modes that exit the Hubble radius earlier (smaller values of $k$), which leads to an overall reddening of the spectrum, converting $n_s$ from the naive expectation for a nearly-massless spectator field ($n_{\rm spec} -1 \simeq 2 \epsilon_* \sim 10^{-3}$) to a value $n_s -1\sim 10^{-2}$ compatible with CMB data, where $n_s (k_*) \equiv 1 + (d \ln {\cal P}_{\cal R}  / d \ln k ) \vert_{k_*}$. The same effect enhances the running of the spectral index, leading to $\alpha_s (k_*)\equiv (d n_s / d \ln k)\vert_{k_*} \sim -5 \times 10^{-3}$, within reach of next-generation experiments. For the fiducial example, we find $n_s= 0.969$ and $\alpha_s =-0.003$ for $N_*=49$.

To contextualize these results, in Fig.~\ref{fig:ns-r}, we compare predictions from this model with current and forecast constraints in the $n_s$-$r$ plane and in the $n_s$-$\alpha_s$ plane. The tensor-to-scalar ratio, $r\simeq 10^{-15}$, is well below the threshold for detection by future experiments. On the other hand, current and future observations of $n_s$ play an important role in constraining the reheating history of the model, with the {\it Planck} 2018 results effectively requiring $N_* > 46$.

Additional constraining power will come from improved measurements of the running of the spectral index $\alpha_s$. Both end points of the range $N_* \in \{ 46.5 , 51.5 \}$, which arise from the residual uncertainty associated with the reheating phase, yield predictions for the $n_s$-$\alpha_s$ plane that are {\it outside} the $2\sigma$ bounds of the expected CMB-S4 constraints, while predictions arising from $N_* < 48.5$ are outside the $2\sigma$ bounds expected from the SPHEREx experiment. Most importantly, combining CMB-S4 with SPHEREx measurements could {\it exclude} this model altogether at the $2\sigma$ level.

We note that, despite the important role of isocurvature perturbations in this model, the exponential decay of isocurvature perturbations ${\cal S}_k$ during the late stages of inflation (see Fig.~\ref{fig:R-and-S}) leads to a negligible primordial isocurvature fraction  $\beta_{\rm iso} \simeq 10^{- 15}$, well below observational constraints on isocurvature in components of the $\Lambda$CDM model \cite{Planck:2018jri}.

Finally, we note that non-Gaussianity in this model is expected to be at most ${\cal O}(1)$. This follows from simple considerations of the power spectrum: the high-$k$ suppression of the curvature perturbation power spectrum (Fig.~\ref{fig:R-and-S}, right panel) implies that the bispectrum should be peaked in the equilateral configuration, with each $k_i \sim k_*$. The equilateral non-Gaussianity can be estimated from standard multifield inflation methods (see, e.g., Ref.~\cite{Kaiser:2012ak}); applied to the scenario under consideration here, this yields $f_{\rm NL}^{\rm equil} \lesssim {\cal O} (1)$. Quantitatively, using the Python package \texttt{PyTransport}~\cite{Mulryne:2016mzv}, we find $f_{\rm NL} ^{\rm equil}=0.48$ for $N_*=49$. \\

\textbf{\textit{Discussion.}} 
In this work, we have discussed a general mechanism by which the model of natural inflation, ostensibly ruled out by current constraints on the tensor-to-scalar ratio $r$, can be brought into agreement with current data. We have presented a proof-of-principle that the tensor-to-scalar ratio can be made exponentially small, $r\sim 10^{-15}$, while retaining excellent agreement between prediction and measurement of the spectral index $n_s$.%
\footnote{
    This complements Ref.~\cite{Montefalcone:2022jfw}, where it was shown that natural inflation coupled to a radiation bath (``warm inflation") can produce $r\sim10^{-10}$.
}
Whereas such tiny values of $r$ are unlikely to be measurable by any future CMB experiments, models such as this one can nonetheless be tested and strongly constrained by considering {\it other} robust observables. In particular, improved measurements of the running of the spectral index, which could come via combination of data from CMB-S4 and SPHEREx, could exclude such models at $>2\sigma$.

These results, complementary to Refs.~\cite{Adshead:2010mc,Hardwick:2018zry,Easther:2021rdg,Bahr-Kalus:2022prj,Martin:2024nlo,Munoz:2016owz}, emphasize that the  ability to test, and even rule out, models of inflation does not lie solely in the hands of the tensor-to-scalar ratio. Rather, the running of the spectral index should serve as a viable test of small-$r$ models. \\

\textbf{\textit{Acknowledgements.}}
The authors thank Richard Easther, Alan H.~Guth, Mikhail Ivanov, Francisco G. Pedro, Wenzer Qin, Michael Toomey, and Vincent Vennin for helpful discussions. Portions of this research were conducted in MIT's Center for Theoretical Physics and supported in part by the U.S.~Department of Energy under Contract No.~DE-SC0012567. E.M. is supported in part by a Discovery Grant from the Natural Sciences and Engineering Research Council of Canada, and by a New Investigator Operating Grant from Research Manitoba.\\

\bibliography{axion-refs}

\end{document}